\documentclass[pra, twocolumn, amscd, amsmath, amssymb,letterpaper,showpacs,biblatex,superscriptaddress]{revtex4-1}
\include{thebibliography}
\usepackage{verbatim}
\usepackage{bbold}
\usepackage{multirow}
\usepackage[pdftex]{graphicx}
\usepackage[pdftex]{epsfig}
\newcommand{\bra}[1]{\left\langle{#1}\right\vert}
\newcommand{\ket}[1]{\left\vert{#1}\right\rangle}

\begin{document}
\title{Comparing algorithms for graph isomorphism using discrete- and continuous-time quantum random walks}
\author{Kenneth Rudinger}
\email{rudinger@wisc.edu} 
\author{John King Gamble}
\affiliation{University of Wisconsin-Madison, Physics Department \\ 1150 University Ave, Madison, Wisconsin 53706, USA}
\author{Eric Bach}
\affiliation{University of Wisconsin-Madison, Computer Sciences Department \\ 1210 W. Dayton St, Madison, Wisconsin 53706, USA}
\author{Mark Friesen}
\author{Robert Joynt}
\author{S. N. Coppersmith}
\email{snc@physics.wisc.edu}
\affiliation{University of Wisconsin-Madison, Physics Department \\ 1150 University Ave, Madison, Wisconsin 53706, USA}

\begin{abstract}
Berry and Wang [Phys. Rev. A {\bf 83}, 042317 (2011)] show numerically that a discrete-time quantum random walk of two noninteracting particles is able to distinguish some non-isomorphic strongly regular graphs from the same family.  Here we analytically demonstrate how it is possible for these walks to distinguish such graphs, while continuous-time quantum walks of two noninteracting particles cannot.  We show analytically and numerically that even single-particle discrete-time quantum random walks can distinguish some strongly regular graphs, though not as many as two-particle noninteracting discrete-time walks.  Additionally, we demonstrate how, given the same quantum random walk, subtle differences in the graph certificate construction algorithm can nontrivially impact the walk's distinguishing power.  We also show that no continuous-time walk of a fixed number of particles can distinguish all strongly regular graphs when used in conjunction with any of the graph certificates we consider.  We extend this constraint to discrete-time walks of fixed numbers of noninteracting particles for one kind of graph certificate; it remains an open question as to whether or not this constraint applies to the other graph certificates we consider.

\end{abstract}

\maketitle

\section{Introduction}
The graph isomorphism problem is considered a good candidate for speedup by a quantum algorithm.  The graph isomorphism (GI) problem is, given two graphs, to determine if they are isomorphic, that is, if one can be mapped to the other via a relabeling of vertices.  To date, the best known classical algorithm has a runtime of $O(c^{\sqrt{N}\log N})$, where $c$ is a constant and $N$ is the number of vertices in the graph \cite{Spielman1996}.  GI is thought to be similar to integer factoring \cite{Shor1994} in that both could reside in the complexity class NP-Intermediate.  Given that there exists a polynomial-time quantum algorithm for integer factoring, it is thought that there may exist a quantum speedup for GI as well.

One class of algorithms that has been explored for GI is that of quantum random walks.  Quantum random walks (QRWs) are of particular algorithmic interest as they are known to offer certain computational advantages over classical random walks \cite{Aharonov1993,Bach2004,Solenov2006}.  There exist problems for which QRW algorithms are known to have faster runtimes than their fastest known classical analogues \cite{Childs2002,Shenvi2003,Ambainis2003,Ambainis2004,Magniez2007,Potocek2009,Reitzner2009}.

Much work has been devoted to developing a quantum algorithm for the graph isomorphism problem.  Several works have examined ways to use continuous-time quantum random walks (CTQRWs) to solve GI \cite{Shiau2005,Gamble2010,Rudinger2012}.  Others have looked to utilize discrete-time quantum random walks (DTQRWs) for GI algorithms \cite{Emms2009,Guo2011}.  While these efforts have helped illuminate what kinds of QRW algorithms might be fruitful for GI, an efficient QRW algorithm for GI has yet to be developed.  Moreover, it is unknown if DTQRWs or CTQRWs are better candidates for an efficient GI algorithm, or if both classes of walks will ultimately be of equal utility.

This question is explored in a recent paper by Berry and Wang \cite{Berry2011}.  Their results indicate that discrete-time quantum random walks may have greater ability to distinguish non-isomorphic strongly regular graphs (a class of graphs which are particularly difficult to distinguish) than continuous-time quantum random walks.  This is based on (1) the ability of a specific discrete-time quantum random walk of two noninteracting particles to distinguish strongly regular graphs (SRGs) that are not distinguished by an algorithm based on a continuous-time quantum random walk of two noninteracting particles \cite{Gamble2010}, and (2) the fact that the dimension of the state space of discrete-time walks on graphs is larger than that of continuous-time walks.  

In this paper, we explain why the two-particle noninteracting walks of Berry and Wang are able to distinguish non-isomorphic SRGs.  To do this, we show analytically how even single-particle DTQRWs are able to distinguish non-isomorphic SRGs.  This demonstrates how two-particle noninteracting DTQRWs can also distinguish SRGs.  We then verify numerically that this distinguishing power of single-particle DTQRWs exists. 

Additionally, we explore the differences between the comparison algorithm used by Berry and Wang and the one used by Gamble \emph{et al.} \cite{Gamble2010}.  Both comparison algorithms rely on the construction of sorted lists of data, or ``graph certificates'', which are used to compare graphs.  Though the respective certificates are constructed in similar manners, we find that, when applied to single-particle discrete-time quantum random walks, there is a non-trivial difference between the distinguishing power of the graph certificates of \cite{Gamble2010} and \cite{Berry2011}.  Thus we demonstrate the importance of using the same kind of graph certificates when attempting to compare the distinguishing power of DTQRWs and CTQRWs.
 
Lastly, we extend a result of Rudinger \emph{et al.} \cite{Rudinger2012}, who showed that there does not exist a CTQRW with a fixed  number of noninteracting particles that can distinguish all SRGs when used in conjunction with the comparison algorithm of Gamble \emph{et al.}  Here we show that this argument holds for DTQRWs as well; no DTQRW with a fixed number of noninteracting particles can distinguish all SRGs when the comparison algorithm of Gamble \emph{et al.} is used.  We also examine the comparison algorithm of Berry and Wang, as well as an algorithm with similarities to both the algorithms of Berry and Wang as well as Gamble \emph{et al}.  We find that no CTQRW of a fixed number of noninteracting particles can distinguish all SRGs when used with these comparison algorithms.  It remains an open question as to whether or not this is true for DTQRWs of a fixed number of noninteracting particles.
 
This paper is organized as follows:  Section~\ref{sec:Back} provides the requisite background regarding graph isomorphism, strongly regular graphs, and quantum random walks.  Section~\ref{sec:Algs} describes in detail the graph comparison procedures of Gamble \emph{et al.}, and Berry and Wang, as well as a procedure which is ``in between'' the two.   In Section~\ref{sec:Analytic}, we analytically show how the single-particle DTQRW has the potential to distinguish non-isomorphic graphs when used with any of the three comparison algorithms we consider.  Section~\ref{sec:Num} provides numerical results for single-particle DTQRWs; we find that all three comparison algorithms numerically distinguish some, but not all, of the tested SRG pairs, and one comparison procedure is not as strong as the other two.  In Section~\ref{sec:Lim}, we show that a DTQRW with a fixed number of noninteracting particles cannot distinguish all SRGs when used with the comparison procedure of Gamble \emph{et al.}; we show the same result holds when the comparison algorithm of Berry and Wang is used with CTQRWs with a fixed number of noninteracting particles.  We are unable to extend this result for Berry and Wang's comparison algorithm when used with a DTQRW of a fixed number of noninteracting particles, indicating a potential difference between DTQRWs and CTQRWs.  We discuss our results in Section~\ref{sec:Disc}.  Lastly, the Appendix provides the explicit formula for a quantity necessary to demonstrate the distinguishing power of the comparison algorithm of Berry and Wang.

\section{Background}
\label{sec:Back}
\subsection{Basic graph definitions}
Here we review the graph-theoretic background necessary to discuss quantum random walks on graphs.  As in \cite{Berry2011,Gamble2010,Rudinger2012}, we consider only simple, undirected graphs.  A graph $G = (V,E)$ is a set of vertices $V$ and edges between those vertices $E$.  The set of vertices $V$ is a set of labels, usually integers, and the edge set $E$ is a set of unordered pairs of vertices.  Two vertices are connected by an edge, and are said to be adjacent or neighboring, if and only if that vertex pair appears in $E$.  A convenient representation of a graph is its adjacency matrix ${\bf A}$:
\begin{equation}
A_{ij} = \begin{cases} 1 & \text{if vertices $i$ and $j$ are adjacent} \\
0 & \text{if vertices $i$ and $j$ are not adjacent.}
\end{cases}
\end{equation}
A graph of $N$ vertices has an $N \times N$ adjacency matrix.  As all graphs we consider are simple and undirected, here ${\bf A}$ is always symmetric, with zeros on the diagonal.

Two graphs are isomorphic if and only if there exists a relabeling of vertices which transforms one graph into the other.  More formally, two graphs represented by adjacency matrices ${\bf A}$ and ${\bf B}$ respectively are said to be isomorphic if and only if there exists a permutation matrix ${\bf P}$ such that ${\bf B} = {\bf P}^{-1}{\bf A P}$.
\subsection{Strongly regular graphs}
The graphs considered by Berry and Wang, as well as Gamble \emph{et al.} and Rudinger \emph{et al.} are called strongly regular graphs (SRGs).  Strongly regular graphs are good test cases for any candidate algorithm for graph isomorphism, as there exist many relatively small SRGs which are difficult to distinguish with the best known classical algorithms.

SRGs are described by four parameters, denoted $(N,d,\lambda,\mu)$.  (Sometimes $k$ is used instead of $d$.)  $N$ is the number of vertices in the graph, and every vertex is adjacent to $d$ other vertices (the graph is $d$-regular).  Every two adjacent vertices share $\lambda$ common neighbors, while every pair of non-adjacent vertices share $\mu$ common neighbors.  SRGs with the same parameters are said to be in the same SRG \emph{family}.  Infinitely many SRG families have more than one non-isomorphic graph \cite{Stones2010}.  Since family parameters are easily computed, it is non-isomorphic graphs in the same family that are difficult to distinguish.  SRG families that contain multiple non-isomorphic graphs are studied in \cite{Berry2011,Gamble2010,Rudinger2012}.

It follows from the definition of SRGs that SRG adjacency matrices satisfy \cite{Godsil2001}:
\begin{equation}
\label{eq:SRG_id}
{\bf A}^2 = (d-\mu) \mathbb{1} + \mu {\bf J} + (\lambda - \mu) {\bf A},
\end{equation}
where $\mathbb{1}$ is the identity matrix and ${\bf J}$ is the matrix of all ones.  Given that ${\bf J}^2 = N {\bf J}$, ${\bf JA} = {\bf AJ} = d{\bf A}$, and $\mathbb{1}$ acts trivially on $\mathbb{1}$, ${\bf J}$, and ${\bf A}$, we see that $\{ \mathbb{1}, {\bf J}, {\bf A} \}$ forms a commutative three-dimensional algebra.  Thus, for any non-negative integer $n$:
\begin{equation}
{\bf A}^n = \alpha_n \mathbb{1} + \beta_n {\bf J} + \gamma_n {\bf A},
\end{equation}
where $\alpha_n$, $\beta_n$, and $\gamma_n$ all depend only on the SRG family parameters and $n$.
\subsection{Discrete-time and continuous-time quantum random walks of noninteracting particles}
Next we discuss how to form quantum random walks on graphs.  For the noninteracting continuous-time model used by \cite{Gamble2010,Rudinger2012}, a Hamiltonian is defined for a graph of $N$ vertices with adjacency matrix ${\bf A}$:
\begin{equation}
\bold{H} = -\sum_{i,j}^N A_{ij} c_i^\dagger c_j,
\end{equation}
where $c_i^\dagger$ and $c_i$ are the creation and annihilation operators, respectively, for a boson or (spinless) fermion on site $i$.
For bosons, they satisfy the commutation relations $[c_i,c_j^\dagger]=\delta_{ij}$ and $[c_i,c_j]=[c_i^\dagger,c_j^\dagger]=0$.  For fermions, they satisfy the anti-commutation relations $\{c_i,c_j^\dagger\}=\delta_{ij}$ and $\{c_i,c_j\} = \{c_i^\dagger,c_j^\dagger\}=0$.  This Hamiltonian is a tight-binding model, where each site corresponds to a vertex in the graph; a particle may move from one vertex to another if the vertices are adjacent.  This Hamiltonian may be used with any number of particles (either bosons or fermions).  A matrix representation of this Hamiltonian for a fixed number of particles is given in \cite{Rudinger2012}.  The evolution operator is then defined in the standard fashion:
\begin{equation}
{\bf U}_C(t) = e^{-i t {\bf H}}, 
\end{equation}
where we have $\hbar =1$ for convenience.  The subscript $C$ denotes that the evolution operator is for a continuous-time walk; discrete-time evolution operators will be denoted with a subscript $D$.

For discrete-time walks, each particle has an auxiliary coin state, which ``points'' to where the particle will next move.  Thus the single-particle walk has basis states of the form $\ket{ij}$, denoting the particle is on site $i$, and its coin is on site $j$, which by definition must be adjacent to $i$.

The noninteracting walk considered by \cite{Berry2011} is based on a single-particle Grover-coined walk.  We follow \cite{Emms2009} and express the evolution operator in expanded form, allowing the coin index to run over all vertices in the graph, but still requiring the particle be adjacent to its coin.  Thus, for an $N$-vertex graph with adjacency matrix ${\bf A}$, the single-particle discrete-time evolution operator is given by
\begin{equation}
{\bf U}_D = {\bf S \Lambda} (\mathbb{1} \otimes {\bf C}) {\bf \Lambda}.
\end{equation}
${\bf \Lambda} \ket{ij} = A_{ij} \ket{ij}$, ensuring that the coin is adjacent to the particle.  ${\bf S}$ is the swap operator; ${\bf S}\ket{ij} = \ket{ji}$.  ${\bf C}$ is the coin operator.  Here it is the Grover coin, which for a $d$-regular graph is  ${\bf C} = -\mathbb{1} + \tfrac{2}{d} {\bf J}$.  This expanded form of ${\bf U}_D$ has the same behavior as in \cite{Berry2011}; we have only introduced $N^2-Nd$ rows of 0, so its dimension is now $N^2 \times N^2$ \cite{Emms2009}.

To advance the walk a discrete number of time steps $t$, ${\bf U}_D$ is applied $t$ times: ${\bf U}_D(t) = ({\bf U}_D)^t$.  Additionally, the evolution operator for $p$ noninteracting particles is just $p$ tensor copies of ${\bf U}_D$: ${\bf U}_{D,p} = ({\bf U}_D)^{\otimes p}$.

\section{COMPARISON ALGORITHMS}
\label{sec:Algs}
Now that we have defined evolution operators for continuous-time and discrete-time walks, we explore how these operators can be used to compare graphs and test for isomorphism.  All methods of comparing graphs that we examine are based on, given an evolution operator, generating a list of numbers sorted by size for each of the graphs, and comparing the respective lists.  We follow \cite{Berry2011} and refer to these sorted lists as graph certificates.

There are multiple graph certificates that can be constructed from a given evolution operator.  In this section, we examine three classes of graph certificates.  First, we give the definition of the certificate which is used by \cite{Gamble2010,Rudinger2012} for CTQRWs.  We then describe a certificate which is ``between'' the certificate of \cite{Gamble2010,Rudinger2012} and the certificate of \cite{Berry2011}.  Lastly, we define the certificate used by \cite{Berry2011} for DTQRWs.  All three certificates can be applied to either continuous-time or discrete-time QRWs.

As the certificates are lists, we denote the three different certificate classes as $L_0$, $L_1$, and $L_2$, respectively.  To indicate that a specific certificate corresponds to a discrete-time walk, a superscript $D$ is included; a superscript $C$ denotes the certificate corresponds to a continuous-time walk.  Additionally, a second superscript is used to indicate that number of particles in the walk which generates the certificate.  For example, $L_2^{D,1}$ would refer to the $L_2$ graph certificate for a discrete-time walk of a single particle.

We call the graph certificate of \cite{Gamble2010} and \cite{Rudinger2012} $L_0$.  It is defined as follows:
\begin{equation}
\label{eq:L_0}
L_0({\bf A},t) = \text{sort}\left(\{|U(t)_{mn}| : \forall m,n \in \dim {\bf U}\}\right).
\end{equation}
Thus, if the possible values the evolution operator elements can take on (up to different phases) are different for the different graphs, then the graphs will be distinguished.  Even if the possible values each element can take on are the same for the different graphs, the graphs will be distinguished if those values have different multiplicities in the two evolution operators.  While this algorithm has been used for CTQRWs, it can in principle be used for DTQRWs as well.  Additionally, we see from its definition that $L_0$ is naturally defined for a walk with any number of particles.

It was proven in \cite{Shiau2005} that $L_0^{C,1}$ certificates could not be used to distinguish non-isomorphic SRGs from the same family, while \cite{Gamble2010} extended this proof for CTQRWs of two noninteracting particles.  However, \cite{Rudinger2012} showed that $L_0$ for CTQRWs with three or more noninteracting particles could distinguish many (but not all) pairs of non isomorphic SRGs from the same family.

The next graph certificate is designed for DTQRWs, but can be used for CTQRWs.  We denote it $L_1$, which for the single-particle DTQRW is defined as
\begin{align}
\label{eq:L_1}
L_1^{D,1}({\bf A}, t) = \text{sort}\Bigg(\Bigg\{\sum_{j=1}^N {|\bra{ij}{\bf U}_{D,1}(t)\ket{kl}|}^2 : \nonumber \\ \forall i,k,l\in \{1\hdots N\}\Bigg\}\Bigg).
\end{align}
Each element of $L_1^{D,1}$ represents the total probability of a particle being on vertex $i$ after $t$ steps, given an initial state $\ket{kl}.$  A natural extension of this method for a two-particle CTQRW is
\begin{align}
L_1^{C,2}({\bf A},t) = \text{sort}\Bigg(\Bigg\{\sum_{j=1}^N {}_B\bra{ij} {\bf U}_{C,2} (t) \ket{kl}_B: \nonumber \\
\forall i,k \in \{1\hdots N\}, \forall l \in \{1 \hdots  k\}\Bigg\}\Bigg),
\end{align}
where $\ket{ij}_B$ denotes the bosonic (symmetrized) state in which there is a boson on site $k$ and a boson on site $l$.  (This method may be used just as well with fermions, with the basis states appropriately anti-symmetrized.)
The extension of $L_1$ for more than two continuous-time particles is given in Section~\ref{subsec:AsympL1L2}.

The final graph certificate we examine is a variant of this method, and is the one utilized by \cite{Berry2011}.  This certificate, denoted $L_2$, is defined for the single-particle DTQRW as follows:
\begin{align}
\label{eq:L_2}
&L_2^{D,1}({\bf A},T) =   \nonumber \\ &\text{sort}\Bigg(\Bigg\{\sum_{t=1}^T\sum_{j=1}^N {|\bra{ij}{\bf U}_{D,1}(t)\ket{kl}|}^2 :\forall i,k,l\in \{1\hdots N\}\Bigg\}\Bigg).
\end{align}
This method sums up the probabilities of a particle being at a particular site at different times, given the same initial state.  For the two-boson CTQRW, this certificate has the form:
\begin{align}
L_2^{C,2}({\bf A},T) = \text{sort}\Bigg(\Bigg\{\sum_{t=1}^T \sum_{j=1}^N {}_B\bra{ij} {\bf U}_{C,2} (t) \ket{kl}_B: \nonumber \\
\forall i,k \in \{1\hdots N\}, \forall l \in \{1 \hdots  k\}\Bigg\}\Bigg).
\end{align}
As with $L_1$, $L_2$ can be extended to any number of discrete-time or continuous-time walkers.

For a two-particle DTQRW, the size of $L_2^{D,2}$ can be quite large, with as many as $N^4d^2$ non-zero elements.  In \cite{Berry2011} the initial states are limited to what Berry and Wang call ``bosonic edge states''.  Given two adjacent vertices $k$ and $l$, a bosonic edge state, denoted by Berry and Wang as $\ket{\beta^+}$, is defined as
\begin{equation}
\ket{\beta^+} = \frac{1}{\sqrt{2}} (\ket{kllk} + \ket{lkkl}).
\end{equation}

Therefore, their graph certificates (which we denote $\widetilde{L_2}^{D,2}$) have at most $\tfrac{N^3d}{2}$ nonzero elements, as a $d$-regular graph with $N$ vertices contains $\tfrac{Nd}{2}$ edges.  The certificate is defined as
\begin{align}
\label{eq:L_2-tilde}
&\widetilde{L_2}^{D,2}({\bf A},T) = \nonumber \\ &\text{sort}\Bigg(\Bigg\{ \sum_{t=1}^T \sum_{j_1,j_2=1}^N {|\bra{i_1 j_1 i_2 j_2}{\bf U}_{D,2}(t)\tfrac{1}{\sqrt{2}}(\ket{kllk}+\ket{lkkl})|}^2 \nonumber \\&:\forall i_1,i_2, \in \{1\hdots N\}, \forall (k,l) \in E\Bigg\}\Bigg),
\end{align}
where $E$ is the edge set of ${\bf A}$.  The physical interpretation of the certificate $\widetilde{L_2}^{D,2}$ is identical to that  of ${L_2}^{D,2}$ except that each initial state is delocalized across an edge in the graph \cite{Berry2011}.
\section{ANALYTIC DEMONSTRATION OF THE DISTINGUISHING POWER OF DTQRWS ON SRGS}
\label{sec:Analytic}
Berry and Wang demonstrated numerically that the noninteracting two-particle DTQRW with the $\widetilde{L_2}$ certificate method could distinguish many SRGs.  In this section, we show analytically why this is possible.  To do so, we first analytically explore the distinguishing power available to single-particle walks using only the $L_0$ and $L_1$ graph certificates.

In general, if a particular kind of certificate will always fail to distinguish two non-isomorphic SRGs, it is because all elements of a certificate, as well as their multiplicities, are functions of SRG family parameters.  This is how we show, or fail to show, the limitations of each certificate considered in this section.

\subsection{Single-particle DTQRW with $L_0$ graph certificate}

Here we show analytically that even single-particle DTQRWs have the potential to distinguish SRGs, using only the $L_0$ graph certificate.  To begin our analysis, we compute an arbitrary element of ${\bf U}_{D,1}$ for $t=1$:

\begin{equation}
\bra{ij}{\bf U}_{D,1}\ket{kl} = A_{ij} A_{kl} \delta_{jk}\left(-\delta_{il} + \frac{2}{d}\right).
\end{equation}
The particle on site $k$ with its coin pointing to site $l$ can move to site $i$ with its coin pointing to site $j$ if and only if $A_{ij}=A_{kl}=\delta_{jk}=1$.  ($A_{ij}=A_{kl}=1$ is required because both the bra and ket must be legal states; a state must always have its coin point to a site that is adjacent to the location of the particle.)  The amplitude of this transition is equal to $\frac{2}{d}$ if $i=l$, and $-1+\frac{2}{d}$ otherwise.  We see that the particle's movement for this single time step is highly restricted, and that all possible evolution operator element values are strictly functions of the family parameters.

Now that we have computed the possible non-zero values for $\bra{ij}U\ket{kl}$, we compute the multiplicities of these values.  If the multiplicities of one value are different for two SRGs in the same SRG family, then the $L_0^{D,1}$ certificates will distinguish the graphs. 

To compute the multiplicities, we follow \cite{Gamble2010,Rudinger2012}.  We denote by $M(x)$ the multiplicity of the value $x$ in the evolution operator, and we recall that the family parameters for an SRG are denoted $(N,d,\lambda,\mu)$.  We find:
\begin{equation}
\label{eq:t1_1}
M\left(\frac{2}{d}\right)=\sum_{ijkl}^N A_{ij} A_{kl} \delta_{jk} (1-\delta_{il}),  
\end{equation}
\begin{equation}
\label{eq:t1_2}
M\left(-1+\frac{2}{d}\right) = \sum_{ijkl}^N A_{ij} A_{kl} \delta_{jk} \delta_{il}. 
\end{equation}

Each of these summands is a product of four terms, each of which corresponds to an identity or adjacency relationship that appears in $\bra{ij}{\bf U}_{D,1}\ket{kl}$.  $1-\delta_{il}$ appears in the first summand, because $\delta_{il}$ is ``turned off" (equals 0) when $\bra{ij}{\bf U}_{D,1}\ket{kl}=\frac{2}{d}$; $\delta_{il}$ appears in the second summand because $\delta_{il}$ is ``turned on" (equals 1) when $\bra{ij}{\bf U}_{D,1}\ket{kl}=-1+\frac{2}{d}$.

The sums given in Eqs.~\eqref{eq:t1_1} and~\eqref{eq:t1_2} can be computed straightforwardly because all the index contractions here are reducible to matrix multiplication and traces.  Additionally, we use the SRG identity of Eq.~\eqref{eq:SRG_id}, and find
\begin{equation}
\label{eq:t1_1a}
M\left(\frac{2}{d}\right)= N \mu(N - d - 1) +  N \lambda d,
\end{equation}
\begin{equation}
\label{eq:t1_2a}
M\left(-1+\frac{2}{d}\right) = N d.
\end{equation}
We see that these multiplicities are functions of the family parameters, so the walk has no distinguishing power when $t=1$.  

Next, we consider later times.  $\bra{ij}{\bf U}_{D,1}^t\ket{kl}$ for $t=2$ and $t=3$ are straightforward to calculate; both the values and degeneracies for these cases are reducible to sums over products of adjacency matrices and traces of adjacency matrices, and therefore can be written as functions of the SRG family parameters.  Therefore, ${\bf U}_{D,1}^2$ and ${\bf U}_{D,1}^3$ cannot distinguish non-isomorphic SRGs of the same family when used to generate $L_0$ certificates.

At $t=4$, all six possible adjacency relations appear ($A_{ij}$, $A_{kl}$, $A_{jk}$, $A_{ik}$, $A_{jl}$ and $A_{il}$):

\begin{align}
\label{eq:t4}
\bra{ij}&U^4\ket{kl}  =  A_{ij}A_{kl}(4d^{-2}(A_{il}-A_{jk}) + 2d^{-1}\times \\
 & (2\delta_{ik} - A_{il} \delta_{ik} - A_{jk} \delta_{ik} + \delta_{jl} - A_{il} \delta_{jl}) + \delta_{ik} \delta_{jl} - \nonumber \\
 & 8d^{-3}((d-\mu)(\delta_{ik}+\delta_{jl}) + (\lambda-\mu)(A_{ik}+A_{jl}) + \nonumber \\
& 2\mu)+16d^{-4} (\delta_{jk} (d-\mu) (\lambda-\mu) + \nonumber \\ 
&A_{jk}(d+(\lambda-\mu)^2-\mu) + (d+\lambda-\mu) \mu)). \nonumber
\end{align}
When all four vertices are connected to each other, $\bra{ij}{\bf U}_{D,1}^4\ket{kl}=\frac{-16\lambda}{d^3} + \frac{16(d+(\lambda-\mu)^2-\mu+(d+\lambda-\mu)\mu)}{d^4}$.  As no other configuration of vertices yields this value for $\bra{ij}{\bf U}_{D,1}^4\ket{kl}$, this value appears in the operator ${\bf U}_{D,1}^4$ $M$ times, where $M$ satisfies

\begin{equation}
\label{eq:t4_example}
M = \sum_{ijkl}^N A_{ij}A_{kl} A_{ik} A_{il} A_{jk} A_{jl}.
\end{equation} 

Computing the sum in Eq.~\eqref{eq:t4_example} requires contracting over four indices, each of which occurs three times.  Such a sum cannot be reduced to be in terms of matrix multiplication and traces.  Ref. \cite{Rudinger2012} showed, in the context of continuous-time noninteracting walks of three particles, that these kinds of sums are functions of the number of shared neighbors belonging to triples of vertices in the graph in question.  This number is, in general, different for different triples of vertices in the same SRG, and therefore not a function of the SRG family parameters.  Therefore, ${\bf U}_{D,1}^4$ has the potential to distinguish non-isomorphic SRGs when used with the $L_0$ method.  In Section~\ref{sec:Num}, we numerically show that this method does indeed distinguish some (but not all) non-isomorphic SRGs.

In contrast, in Refs. \cite{Shiau2005, Gamble2010} it is proved that single-particle and noninteracting two-particle CTQRWs with the $L_0$ graph certificate method cannot distinguish any non-isomorphic SRG pair from the same family.  Several key differences between DTQRWs and CTQRWs become apparent.  First, an element of the single-particle continuous-time walk evolution operator is indexed by up to $2$ vertices (as the particle is on $1$ vertex in an initial state, $1$ vertex in the final state, and the sets of final and initial vertices need not overlap).  However, an element of the corresponding single-particle discrete-time walk evolution operator is indexed by up to $4$ vertices, because each particle corresponds to a vertex, as does each particle's coin.  Additionally, in the continuous-time walk, the value of an evolution operator element does not depend on whether or not two vertices which are both in the final or initial state are adjacent.  Therefore, such adjacency relations are not considered when performing the appropriate multiplicity sum.  However, because a particle must always be adjacent to its coin, the adjacency relation between a particle and its coin is always included in the discrete-time sum used to compute element multiplicity.  The presence of these additional adjacency relations gives the single-particle DTQRW with the $L_0$ comparison protocol the potential to distinguish non-isomorphic SRGs.

\subsection{Single-particle DTQRW with $L_1$ graph certificate}
\label{subsec:L_1}
Here we show that the $L_1$ method, when used with the single-particle DTQRW, also has the potential to distinguish non-isomorphic SRGs from the same family.  To start, we compute an element of $\sum_{j=1}^N|\bra{ij}{\bf U}_{D,1} \ket{kl}|^2$ for $t=1$.  
\begin{equation}
\sum_{j=1}^N|\bra{ij}{\bf U}_{D,1} \ket{kl}|^2 = A_{ik} A_{kl}\left(\delta_{il}\left(1-\tfrac{4}{d}\right)+\tfrac{4}{d^2}\right)
\end{equation}
Possible non-zero values for $\sum_{j=1}^N|\bra{ij}{\bf U}_{D,1} \ket{kl}|^2$ are $\tfrac{4}{d^2}$ and $1-\tfrac{4}{d}+\tfrac{4}{d^2}$, which are both functions of the family parameters.  We may compute their multiplicities, and find that they are also functions of the family parameters.  Thus, $L_1^{D,1}(t=1)$ graph certificates cannot distinguish non-isomorphic SRGs from the same family.

Similarly, it may be shown that at $t=2$ for the single-particle DTQRW, all elements of $L_1$ are functions of the family parameters, as well as their multiplicities.  Therefore,  $L_1^{D,1}(t=2)$ certificates also cannot distinguish non-isomorphic SRGs from the same family.

However, at $t=3$, the values of the elements of $L_1$ for the single-particle DTQRW at $t=3$ are \emph{not} functions of the family parameters.  The Appendix provides the value of $\sum_{j=1}^N |\bra{ij} {\bf U}_{D,1}(3) \ket{kl}|^2$, which has the form:
\begin{equation}
\label{eq:L_1-sum}
\sum_{j=1}^N |\bra{ij} {\bf U}_{D,1}(3) \ket{kl}|^2 = g(i,k,l) + h(i,k,l)\sum_{j=1}^N A_{ij}A_{jl} A_{jk},
\end{equation}
where $g(i,k,l)$ and $h(i,k,l)$ are functions of the family parameters.  However, $\sum_{j=1}^N A_{ij}A_{jl} A_{jk}$ is not a function strictly of family parameters, for the same reason that Eq.~\eqref{eq:t4_example} is not.  Thus we do not even need to examine the multiplicities of different values of $\sum_{j=1}^N |\bra{ij} {\bf U}_{D,1}(3) \ket{kl}|^2$, as this sum takes on values that are not functions of the SRG family parameters.  Therefore, the single-particle $L_1$ method at $t=3$ can potentially distinguish non-isomorphic SRGs from the same family.  In Section~\ref{sec:Num}, we numerically show that this method can distinguish some non-isomorphic graph pairs from the same SRG family.

However, this is not true for two-particle noninteracting CTQRWs.  Using the methods of \cite{Gamble2010,Rudinger2012}, it can be shown that, when the appropriately symmetrized or anti-symmetrized states are used with the continuous-time evolution operator, all possible values and corresponding multiplicities of $\sum_{j=1}^N |\bra{ij} {\bf U}_{C,2}(t) \ket{kl}|^2$ are functions of the family parameters and the time $t$.  Thus, we see that the single-particle DTQRW $L_1$ method is strictly stronger than the two-particle noninteracting CTQRW $L_1$ method for distinguishing SRGs. 

\subsection{Single-particle DTQRW with $L_2$ graph certificate}

Now we examine $L_2^{D,1}(T)$ analytically for varying values of $T$.  Recall that elements of $L_2^{D,1}(T)$ are sums of elements from $L_1^{D,1}(t)$ with $t$ running from $1$ to $T$.  By inspection of Eqs.~\eqref{eq:L_1} and~\eqref{eq:L_2}, we see that for a final time of $T=1$, $L_2$ is the same as $L_1(t=1)$, and will have no distinguishing power.  Similarly, for $T=2$, each element of $L_2$ will be a sum of a term from $L_1(t=1)$ and $L_1(t=2)$.  As the value and multiplicity of each of those terms is strictly a function of SRG family parameters, the corresponding values and multiplicities of each element of $L_2(T=2)$ will be a function of SRG family parameters.  Thus $L_2^{D,1}(T=2)$ certificates cannot distinguish non-isomorphic SRGs from the same family.

However, we know that $L_1^{D,1}(t=3)$ certificates have the potential to distinguish SRGs.  Thus, $L_2^{D,1}(T=3)$ certificates also have the potential to distinguish SRGs from the same family, as the elements of a $L_2^{D,1}(T=3)$ certificate are sums of terms which include elements of a $L_1^{D,1}(t=3)$ certificate, which we have shown to not be functions of family parameters.

Two-particle noninteracting CTQRWs with the $L_2$ method may be similarly analyzed.  Because the $L_1$ method fails to distinguish SRGs with noninteracting two-particle CTQRWs for all values of $t$, these walks with the $L_2$ method will also be unable to distinguish SRGs.  Thus we analytically see that for all three comparison methods contemplated, the single-particle DTQRW has the potential to distinguish non-isomorphic SRGs from the same family, while two-particle noninteracting CTQRWs cannot.
\section{NUMERICAL RESULTS}
\label{sec:Num}

\begin{table*}[tbh!]
\begin{tabular}{| c | c | c || c | c | c | c | c | c | c |}
\hline
&  &  & \multicolumn{6}{c}{Number of Undistinguished Pairs} & \\
\hline
SRG Family & Number & Comparisons & $L_0^{D,1} (t=4)$ &$ L_0^{D,1} (t=2N) $&$ L_1^{D,1} (t=3) $ &$L_1^{D,1} (t=4)$ & $L_2^{D,1}(T=3)$ & $L_2^{D,1}(T=4)$ & $L_2^{D,1}(T=2N)$\\ 
($N$,$d$,$\lambda$,$\mu$) & of Graphs & & &  & & &  &  &\\
\hline
(16,6,2,2) & 2 & 1 & 1 & 1 & 1 & 1 & 1 & 1 & 1\\
\hline
(25,12,5,6) & 15 & 105 & 11 & 11 & 0 & 0 & 0 & 0 & 0\\
\hline
(26,10,3,4) & 10 & 45 & 3 & 3 &1 & 1 & 1 & 1 & 1\\
\hline
\end{tabular}
\caption{ \label{tab:results1} Numerical results for graph isomorphism testing using single-particle DTQRWs with varying comparison algorithms with various times.  The different graph certificates ($L_0^{D,1}$, $L_1^{D,1}$, and $L_2^{D,1}$) are defined in Eqs.~\eqref{eq:L_0},~\eqref{eq:L_1}, and~\eqref{eq:L_2}.  We note that all three algorithms have significant, but not universal, distinguishing power on SRGs.  Additionally, this distinguishing power saturates at the minimum time at which each algorithm can potentially distinguish SRGs.  Lastly, we see that $L_1^{D,1}$ and $L_2^{D,1}$ are more powerful than $L_0^{D,1}$, but $L_2^{D,1}$ is possibly no more powerful than $L_1^{D,1}$.}
\end{table*}

We now test the three comparison methods with the single-particle DTQRW on three families of SRGs.  For the DTQRWs, all three algorithms are tested at the minimum time they can potentially distinguish SRGs ($t=4$ for $L_0^{D,1}$, $t=3$ for $L_1^{D,1}$, and $T=3$ for $L_2^{D,1}$).  We additionally test $L_0^{D,1}$ at $t=2N$, $L_1^{D,1}$ at $t=4$, and $L_2^{D,1}$ at $T=4$ and $T=2N$.  ($T=2N$ is the total time used by Berry and Wang for their tests \cite{Berry2011}.)  Our results are given in Table~\ref{tab:results1}.

Each algorithm distinguishes some, but not all, non-isomorphic SRGs from the same family.  Also, no single-particle DTQRW algorithm that we test is able to distinguish the pair of SRGs in $(16,6,2,2)$, which is in contrast to the results of Berry and Wang, who successfully distinguished these graphs using the two-particle DTQRW with $\widetilde{L_2}^{D,2}$ certificates.  Additionally, the various algorithms we test have significant distinguishing power on the other two SRG families we examine.  We note that the distinguishing power for each algorithm saturates at the minimum time required to allow for potentially distinguishing SRGs.  This could be because SRGs are distance-regular with diameter 2, so allowing the particles to walk for longer does not actually result in capturing more of the graph structure.

The distinguishing power seems to saturate with $L_1^{D,1}(t=3)$, for higher times, either with $L_1^{D,1}$ or $L_2^{D,1}$, no further distinguishing power is obtained.  For $(25,12,5,6)$, $L_1^{D,1}(t=3)$ distinguishes all possible graph pairs.  For $(26,10,3,4)$, there is one graph pair that cannot be distinguished by $L_1^{D,1}(t=3)$, $L_1^{D,1}(t=4)$, or $L_2^{D,1}$ with any time we test.  Therefore, it is possible that $L_2^{D,1}$ offers no more distinguishing power than $L_1^{D,1}$.  More generally, it is possible that $L_2^{D,p}$ yields no more distinguishing power than $L_1^{D,p}$.  Additionally, it is possible that there exists a certain time (and perhaps relatively small time) beyond which no additional distinguishing power is obtained.  As the complexity of computing $L_1^{D,p}$ and $L_2^{D,p}$ increase with the number time steps used, it could useful to know if there is no additional information to be gained by increasing the number of time steps.

\section{ASYMPTOTIC LIMITS FOR DTQRWS AND CTQRWS}
\label{sec:Lim}
We have shown how, by all three comparison algorithms considered, the single-particle DTQRW has more distinguishing power on SRGs than noninteracting two-particle CTQRWs (which cannot distinguish any non-isomorphic pair of SRGs from the same family).  However, as the distinguishing power of the single-particle DTQRW is not universal on SRGs, we can contemplate how to increase this distinguishing power.  One obvious method is to add additional particles to the walk.  Indeed, with a two-particle noninteracting DTQRW, Berry and Wang distinguished SRG pairs that are not distinguished by $L_2^{D,1}$ certificates for $T=3,4$ and $2N$ \cite{Berry2011}.  Thus it is interesting to characterize the asymptotic distinguishing power of many-body QRWs.  In this section, we show that no CTQRW of a fixed number of noninteracting particles can distinguish all SRGs when used with either $L_0$, $L_1$, or $L_2$ certificates.  We also show this result holds for DTQRWs of any fixed number of noninteracting particles when used to generate $L_0$ certificates; we are unable to extend this result for such walks with $L_1$ and $L_2$ certificates.

\subsection{Asymptotic behavior for $L_0$}

It was demonstrated in \cite{Rudinger2012} that increasing the number of noninteracting particles in a CTQRW could significantly increase distinguishing power of the walk when used with the $L_0$ method.  However, it was also shown in \cite{Rudinger2012}  that no CTQRW with a fixed number of noninteracting particles could distinguish all SRGs.  This was done by showing that, as graph size increases, the number of unique SRGs with the same family parameters is super-exponentially large in graph size \cite{Stones2010,McKay2005}, while for a fixed particle number $p$, the number of unique graphs that the $p$-particle CTQRW can distinguish with the $L_0$ method is polynomial in graph size.

We show here that this limitation of $L_0$ certificates applies to DTQRWs as well.  Let us consider the $p$-particle noninteracting DTQRW, and let us denote by $X_p$ the maximum number of unique values that $\bra{i_1,j_i \hdots i_p, j_p} {\bf U}_{D,p}^t \ket{k_1,l_1 \hdots k_p, l_p}$ can take on.  $X_p$ is \emph{not} a function of the size of the graph; the value of a DTQRW evolution operator element for an SRG is determined only by the configuration of up to $4p$ vertices in that SRG.  (For further details, see \cite{Rudinger2012}.)  

Additionally, the number of elements in ${\bf U}_{D,p}$ is $N^{4p}$, where $N$ is the number of vertices in the graph.  Thus, for an SRG family with graphs of $N$ vertices, the $L_0$ method generates certificates of length $N^{4p}$, where each element in the certificate can take on a maximum number of $X_p$ different values.  Thus, given an SRG family with graph size $N$,  the maximum number of unique $L_0$ graph certificates the noninteracting $p$-particle DTQRW can generate is equivalent to the number of ways one can put $N^{4p}$ indistinguishable balls into $X_p$ distinguishable bins, or ${X_p + N^{4p} -1} \choose {X_p -1}$ \cite{Tucker2004}.

As $X_p$ is constant in $N$, this quantity is polynomial in $N$.  Thus, just like noninteracting CTQRWs with fixed particle number, DTQRWs with fixed particle number cannot distinguish all SRGs with the $L_0$ graph certificate method.  Thus, the $L_0$ method for noninteracting $p$-particle walks \emph{cannot} yield a universal GI algorithm for SRGs, whether or not the walks are continuous-time or discrete-time.

\subsection{Asymptotic behavior for $L_1$ and $L_2$}
\label{subsec:AsympL1L2}
For discrete-time walks, the $L_1$ graph certificate does not necessarily have the same limitations as $L_0$.  This is because the values of the elements of discrete-time $L_1$ certificates are \emph{not} functions of SRG family parameters, as demonstrated by Eq.~\eqref{eq:L_1-sum}.  For a fixed number of discrete-time walkers, the number of unique values that $\sum_{j_1 \hdots j_p=1}^N |\bra{i_1 j_1 \hdots i_p j_p}{\bf U}^t_{D,p} \ket{k_1 l_1 \hdots k_p l_p}|^2$ can take on is not a function only of particle number.  In principle, this quantity may be a function of the graph size, in which case the number of unique $L_1$ graph certificates for a fixed particle number and SRG family may be super-exponential in graph size.  Thus, the number of unique SRGs in a family with graph size $N$ is \emph{not} guaranteed to be larger than the number of unique discrete-time $L_1$ graph certificates that can be generated for walks of $p$ noninteracting particles on graphs of $N$ vertices.  Thus, the proof of the limitations of the $L_0$ method fails to translate to the $L_1$ method for DTQRWs.  Whether there exists such a limitation is therefore still an open question.

Because DTQRW $L_2$ graph certificates are also affected by sums of the form shown in Eq.~\eqref{eq:L_1-sum}, we conclude that, for DTQRWs, the proof of the limitations of $L_0$ cannot be applied to $L_2$, just as it cannot be applied to $L_1$.  Therefore, for DTQRWs with $L_0$ and $L_1$ graph certificates, it is possible that there exists a fixed $p$ such that the $p$-particle noninteracting walk can distinguish all SRGs.

We will now show that, using the same method of proof from the previous subsection, no CTQRW of a fixed number of noninteracting particles can distinguish all SRGs, when used in conjunction either the $L_1$ or $L_2$ graph certificates.  We will only examine bosonic walks; the results for fermionic walks can be proved in the same manner.

We begin by stating the definition of $L_1$ for $p$-boson CTQRWs.  There are multiple ways in which we can extend $L_1$ for more than two particles in a CTQRW, as we can choose the number of particles in the final state whose probability distribution we wish to measure.  For simplicity, here we sum over one particle.  However, our proof holds if we sum over more than one particle as well.

We examine the certificate $L_1^{C,p}(t)$:

\begin{align}
L_1^{C,p}(t)= \text{sort}\{\sum_{i_1}^N |_B\bra{i_1 \hdots i_p} {\bf U}_{C,p}(t) \ket{j_1 \hdots j_p}_B|^2: \nonumber \\
 1 \leq  i_2 \leq i_3 \hdots \leq i_p \leq N, 1 \leq  j_1 \leq j_2 \hdots \leq j_p \leq N\}.
\end{align}

It has been shown \cite{Gamble2010, Rudinger2012} that for SRGs, elements of the noninteracting $p$-boson CTQRW evolution operator have the following simple form:
\begin{align}
_B\bra{i_1 \hdots i_p} {\bf U}_{C,p}(t) \ket{j_1 \hdots j_p}_B = \nonumber \\ _B\bra{i_1 \hdots i_p} (\alpha \mathbb{1} + \beta {\bf J} + \gamma {\bf A})^{\otimes p} \ket{j_1 \hdots j_p}_B
\end{align}
where $\alpha$, $\beta$, and $\gamma$ are all functions of the family parameters and the time $t$.  (The fermionic walks have the same form, except the states are anti-symmetrized, and $\alpha$, $\beta$, and $\gamma$ are replaced by their respective complex conjugates.)

Therefore, $ _B\bra{i_1 \hdots i_p} {\bf U}_{C,p}(t) \ket{j_1 \hdots j_p}_B$ is a function of family parameters, $t$, and the binary relationships $A_{xy}$ and $\delta_{xy}$ for $x \in \{i_1 \hdots i_p\}$ and $y \in \{j_1 \hdots j_p\}$.  Moreover, this quantity is a sum of terms where each term is a product of up to $p$ binary relationships and various powers of $\alpha$, $\beta$, $\gamma$.  Each term does not contain any more than one instance of any one of the $2p$ indices.

Hence $|_B\bra{i_1 \hdots i_p} {\bf U}_{C,p}(t) \ket{j_1 \hdots j_p}_B|^2$ contains no term which includes more than two instances of the same index.  Thus, $\sum_{i_1}^N |_B\bra{i_1 \hdots i_p} {\bf U}_{C,p}(t) \ket{j_1 \hdots j_p}_B|^2$ is only a function of family parameters and the binary relationships $A_{xy}$ and $\delta_{xy}$ for $x \in \{i_2 \hdots i_p\}$ and  $y \in \{j_1 \hdots j_p\}$.  For this quantity to \emph{not} be a function of family parameters, the index $i_1$ would have to appear three times in a single term in $|_B\bra{i_1 \hdots i_p} {\bf U}_{C,p}(t) \ket{j_1 \hdots j_p}_B|^2$, as explained in \cite{Rudinger2012}.

The value of each unique element in $L_1^{C,p}(t)$ is determined by relationships between the $2p-1$ indices $\{i_2 \hdots i_p, j_1 \hdots j_p\}$.  Following \cite{Rudinger2012}, it can be shown that the number of unique elements $L_1^{C,p}(t)$ can contain is bounded above by $2^{p^2+O(p\log p)}$.  Additionally, the length of $L_1^{C,p}(t)$ is ${{N + p - 1} \choose p} {{N+p - 2 } \choose {p-1}}$.  Following the same argument from the end of the previous subsection and from Section IV C of \cite{Rudinger2012}, the number of unique graph certificates this process can generate for a fixed particle number $p$ is polynomial in graph size $N$.  
Thus a CTQRW used with the $L_1$ method and a fixed number of noninteracting particles cannot be universal for SRGs, as there will exist more SRGs than unique $L_1^{C,p}(t)$ certificates.

This analysis extends to $L_2^{C,p}(T)$.  The possible values elements of $L_2^{C,p}(T)$ will in general be different from the possible values of $L_1^{C,p}(t)$.  However, the number of possible values elements of $L_2^{C,p}(T)$ can take on will still be bounded above by $2^{p^2+O(p\log p)}$.  Additionally, the lengths of the $L_2^{C,p}(T)$ and $L_1^{C,p}(t)$ certificates are the same.  Thus the number of SRGs of size $N$ in a single family that $L_2^{C,p}(T)$ certificates can distinguish will be polynomial in $N$.  We conclude that computing $L_2^{C,p}(T)$ certificates for a fixed number $p$ of noninteracting particles cannot distinguish all SRGs.

\section{SUMMARY}
\label{sec:Disc}
We have shown how single-particle discrete-time quantum random walks can distinguish many non-isomorphic strongly regular graphs.  These results are proven with techniques used to analyze the distinguishing power of noninteracting continuous-time quantum random walks \cite{Shiau2005,Gamble2010,Rudinger2012}.   These results regarding single-particle DTQRWs in turn explain the results of Berry and Wang \cite{Berry2011}, who numerically found that two-particle discrete-time walks could distinguish many strongly regular graphs.\\
\indent Additionally, we have examined a proposal of \cite{Berry2011}, that DTQRWs have more distinguishing power than CTQRWs.  To evaluate this proposal, we have found that it is important to consider not just the kind of QRW in question, but the method in which the graph certificate is constructed.  We have considered three related graph certificate construction methods, which we have dubbed $L_0$, $L_1$, and $L_2$.\\
\indent We have found that single-particle DTQRWs used with the $L_0$ method can distinguish many SRGs, in contrast to single-particle and noninteracting two-particle CTQRWs, which, when used with the $L_0$ method, cannot distinguish any SRGs from the same family, as proven in \cite{Shiau2005,Gamble2010}.  However, we have also extended the results of \cite{Rudinger2012}, which showed that there does not exist a fixed particle number $p$ such that a noninteracting $p$-particle CTQRW with the $L_0$ can distinguish all SRGs.  Here we have shown this limitation to hold true for noninteracting $p$-particle DTQRWs as well.\\
\indent Lastly, we have shown that this limitation holds for CTQRWs when the $L_1$ and $L_2$ certificate methods are considered.  There does not exist a fixed number $p$ such that a noninteracting $p$-particle CTQRW with either $L_1$ or $L_2$ certificates can distinguish all SRGs.  However, it remains an open question as to whether or not these limitations of $L_1$ and $L_2$ apply to DTQRWs.  Thus it is possible that there exists a noninteracting $p$-particle DTQRW such that $L_1$ or $L_2$ certificates can distinguish all SRGs.  This would demonstrate a nontrivial difference in distinguishing power between continuous-time and discrete-time noninteracting walks.

\section{Acknowledgements}
This work was supported in part by ARO, DOD (W911NF-09-1-0439) and NSF (CCR-0635355).  We thank Jingbo Wang and Adam Frees for helpful discussions.\\

\section{Appendix}
Eq.~\eqref{eq:L_1-sum} in Section~\ref{subsec:L_1} demonstrates that $L_1^{D,1}$ certificates can distinguish non-isomorphic SRGs of the same family.  To show that Eq.~\eqref{eq:L_1-sum} is of the correct form and cannot be a function of family parameters, we provide here the value of $\sum_{j=1}^N |\bra{ij} {\bf U}_{D,1}(3) \ket{kl}|^2$:

\begin{align}
\label{eq:L1_3}
& \sum_{j=1}^N |\bra{ij} {\bf U}_{D,1}(3) \ket{kl}|^2= \\ 
& \sum_{j=1}^N A_{ij}A_{kl} ( (A_{ik} + A_{jl}) \times \nonumber \\ 
& \left(\tfrac{-16}{d^4} (A_{ik}+A_{jl}) -\tfrac{16}{d^3} \delta_{il} + \tfrac{8}{d^2} \delta_{il} \delta_{jk} - \tfrac{64}{d^5} A^2_{jk} \right) +\nonumber \\
& \delta_{il} \left(\tfrac{4}{d^2} - \tfrac{4}{d} \delta_{jk} + \tfrac{32}{d^4}A^2_{jk} +\delta_{jk} \left(1 - \tfrac{16}{d^3}A^2_{jk}\right)\right) + \tfrac{64}{d^5} \left(A^2_{jk}\right)^2) \nonumber 
\end{align}

We recall the SRG identity of Eq.~\eqref{eq:SRG_id}, and see that\\ $A^2_{jk} = (d-\mu)\delta_{jk} + \mu + (\lambda-\mu)A_{jk}$.  Thus by inspection of Eq.~\eqref{eq:L1_3}, we see that $\sum_{j=1}^N |\bra{ij} {\bf U}(3) \ket{kl}|^2$ contains a term proportional to $A_{kl}\sum_{j=1}^N A_{ij}A_{jl} A_{jk}$, corroborating Eq.~\eqref{eq:L_1-sum}.  Therefore, no element of $L_1^{D,1}$ can be a function only of SRG family parameters, as explained in Section~\ref{subsec:L_1}.
\bibliography{discretetime}

\end{document}